\documentclass{nature}

\usepackage[T1]{fontenc}
\usepackage{textcomp}
\usepackage{amssymb}
\usepackage{amsmath}

\usepackage{gensymb}
\usepackage{caption}
\usepackage{bm}
\usepackage{lineno}
\usepackage{multibib}
\usepackage{xcolor}
\usepackage{hyperref}
\usepackage{afterpage}
\usepackage{epstopdf} 

\newcites{main}{References}
\newcites{methods}{References for Methods}

\usepackage{graphicx}
\makeatletter
\let\saved@includegraphics\includegraphics
\AtBeginDocument{\let\includegraphics\saved@includegraphics}
\renewenvironment*{figure}{\@float{figure}}{\end@float}
\makeatother

\title{A rapidly-changing jet orientation in the stellar-mass black hole V404 Cygni}

\author{James C.\ A.\ Miller-Jones$^{1}$, Alexandra J.\ Tetarenko$^{2,3}$, Gregory R.\ Sivakoff$^{2}$, Matthew J.\ Middleton$^4$, Diego Altamirano$^{4}$, Gemma E.\ Anderson$^{1}$, Tomaso M.\ Belloni$^{5}$, Rob P.\ Fender$^{6}$, Peter G.\ Jonker$^{7,8}$, Elmar G.\ K\"ording$^{8}$, Hans A.\ Krimm$^{9,10}$, Dipankar Maitra$^{11}$, Sera Markoff$^{12,13}$, Simone Migliari$^{14,15}$, Kunal P.\ Mooley$^{6,16,17}$, Michael P.\ Rupen$^{18}$, David M.\ Russell$^{19}$, Thomas D.\ Russell$^{12}$, Craig L.\ Sarazin$^{20}$, Roberto Soria$^{21,1,22}$, Valeriu Tudose$^{23}$}

\begin{document}

\maketitle

\begin{affiliations}
 \item International Centre for Radio Astronomy Research -- Curtin University, GPO Box U1987, Perth, WA 6845, Australia 
 \item Department of Physics, University of Alberta, 4-181 CCIS, Edmonton, AB T6G 2E1, Canada
 \item East Asian Observatory, 660 N. A'ohoku Place, University Park, Hilo, Hawaii 96720, USA
 \item School of Physics \& Astronomy, University of Southampton, Southampton SO17 1BJ, United Kingdom
 \item{INAF -- Osservatorio Astronomico di Brera, Via E. Bianchi 46, I-23807 Merate (LC), Italy}
 \item{Astrophysics, Department of Physics, University of Oxford, Keble Road, Oxford OX1 3RH, UK}
 \item{SRON, Netherlands Institute for Space Research, Sorbonnelaan 2, 3584 CA Utrecht, the Netherlands}
 \item{Department of Astrophysics/IMAPP, Radboud University, Nijmegen, PO Box 9010, 6500 GL Nijmegen, the Netherlands}
\item{Universities Space Research Association, 7178 Columbia Gateway Dr, Columbia, MD 21046, USA} 
\item{National Science Foundation, 2415 Eisenhower Ave, Alexandria, VA 22314, USA} 
\item{Department of Physics \& Astronomy, Wheaton College, Norton, MA 02766, USA}
 \item{Anton Pannekoek Institute for Astronomy, University of Amsterdam, Science Park 904, 1098 XH Amsterdam, the Netherlands}
 \item{Gravitation Astroparticle Physics Amsterdam (GRAPPA) Institute, Science Park 904, 1098 XH Amsterdam, the Netherlands}
 \item{ESAC/ESA, XMM-Newton Science Operations Centre, Camino Bajo del Castillo s/n, Urb. Villafranca del Castillo, 28692, Villanueva de la Ca\~nada, Madrid, Spain}
\item{Institute of Cosmos Sciences, University of Barcelona, Mart\'i i Franqu\`es 1, 08028 Barcelona, Spain}
 \item{NRAO, P.O.\ Box O, Socorro, NM 87801, USA}
 \item{Caltech, 1200 E. California Blvd., MC 249-17, Pasadena, CA 91125, USA}
 \item{Herzberg Astronomy and Astrophysics Research Centre, 717 White Lake Road, Penticton, BC V2A 6J9, Canada}
 \item{New York University Abu Dhabi, P.O.\ Box 129188, Abu Dhabi, United Arab Emirates}
 \item{Department of Astronomy, University of Virginia, 530 McCormick Road, Charlottesville, VA, 22903, USA}
 \item{School of Astronomy and Space Sciences, University of the Chinese Academy of Sciences, Beijing 100049, China}
 \item{Sydney Institute for Astronomy, School of Physics A28, The University of Sydney, Sydney, NSW 2006, Australia}
 \item{Institute for Space Sciences, Atomistilor 409, PO Box MG-23, 077125 Bucharest-Magurele, Romania}
\end{affiliations}


\begin{abstract}
Powerful relativistic jets are one of the main ways in which accreting black holes provide kinetic feedback to their surroundings. Jets launched from or redirected by the accretion flow that powers them should be affected by the dynamics of the flow, which in accreting stellar-mass black holes has shown increasing evidence for precession\citemain{ingram16} due to frame dragging effects that occur when the black hole spin axis is misaligned with the orbital plane of its companion star\citemain{lense18}. 
 Recently, theoretical simulations have suggested that the jets can exert an additional torque on the accretion flow\citemain{liska18}, although the full interplay between the dynamics of the accretion flow and the launching of the jets is not yet understood. Here we report a rapidly changing jet orientation on a timescale of minutes to hours in the black hole X-ray binary V404 Cygni, detected with very long baseline interferometry during the peak of its 2015 outburst. We show that this can be modelled as Lense-Thirring precession of a vertically-extended slim disk that arises from the super-Eddington accretion rate\citemain{motta17}. Our findings suggest that the dynamics of the precessing inner accretion disk could play a role in either directly launching or redirecting the jets within the inner few hundred gravitational radii. Similar dynamics should be expected in any strongly-accreting black hole whose spin is misaligned with the inflowing gas, both affecting the observational characteristics of the jets, and distributing the black hole feedback more uniformly over the surrounding environment\citemain{vernaleo06,falceta-goncalves10}.
\end{abstract}

During the 2015 outburst\citemain{rodriguez15} of the black hole X-ray binary system V404 Cygni\citemain{shahbaz94}, we conducted high-angular resolution radio monitoring with the Very Long Baseline Array (VLBA).
Our observations (Extended Data Table 1) spatially resolved the jets in this system, on size scales of up to 5 milliarcseconds (12 a.u.\ at the known distance of $2.39\pm0.14$\,kpc\citemain{miller-jones09}; see examples in Figure~1). These jets evolved in both morphology and brightness on timescales of minutes.

The orientation of the jets on the plane of the sky varied between epochs, ranging between $-30.6$\degree\ and $+5.6$\degree\ east of north (Figure~1, 2, and Extended Data Table~2).  This range encompasses the orientation inferred from the position angle of the linearly-polarised radio emission\citemain{corbel00} measured during the 1989 outburst ($-16\pm6$\degree\ east of north; we state all uncertainties at 68\% confidence)\citemain{han92}. Moreover, during a period of intense radio and sub-millimetre flaring on June 22nd\citemain{tetarenko17}, we observed multiple ejection events spanning a similar range of orientations over a single four-hour observation (Figure~1), implying extremely rapid changes in the jet axis.

The time-resolved images from June 22nd (see Supplementary Video) show a series of ballistically-moving ejecta that persist for tens of minutes before fading below the detection threshold of $\approx 10$\,mJy.
The radio emission is dominated by a stationary core that is always present, allowing us to perform relative astrometry on the ejecta.
The ejecta appear on both sides of the core, with proper motions ranging from 4.3 to 46.2 milliarcseconds (mas) day$^{-1}$ (0.06--0.64$c$ in projection; Figure~3), at position angles between $-28.6$\degree\ and $-0.23$\degree\ east of north on the plane of the sky (Extended Data Figures 1--4; Extended Data Table 3). 

Under the (standard) assumption of intrinsic symmetry, then with the known distance\citemain{miller-jones09} we can use the measured proper motions of corresponding pairs of approaching and receding ejecta to determine $\theta$, the inclination angle to the line of sight, as well as the dimensionless jet speed $\beta = v/c$ (see Methods). We identify three likely pairs of ejecta with consistent position angles and ejection times (denoted N2/S2, N3/S3 and N6/S6; see Figure 3 and Extended Data Figures 1--3), although since their flux density evolution cannot be fully explained by Doppler boosting of intrinsically symmetric jets (see Methods), the assumption of symmetry remains unverified. From these three pairs we determine ($\beta=0.32\pm0.02$, $\theta=40.6\pm2.4$\degree), ($\beta=0.35\pm0.01$, $\theta=32.5\pm1.6$\degree), and ($\beta=0.48\pm0.01$, $\theta=14.0\pm0.8$\degree), respectively (Figure~4). In all three cases the northern component is the faster-moving, and must therefore be the approaching component.
For unpaired ejecta, we can use the known distance to solve for $\beta\cos\theta$, subject to an assumption on whether the components are approaching or receding (Figure~4).  Again, we find that the jet speed or inclination angle, or both, must vary between ejection events.

The most natural interpretation for changes in jet orientation is precession, as best studied in the persistent X-ray binary SS\,433.
However, each individual jet component only samples the orientation of the jet axis at the time of ejection. With only twelve discrete components on June 22, we do not have sufficient sampling to determine whether the precession is regular.  Our best constraint on the precession period comes from the $\sim30$\degree\ swing in position angle between ejecta pairs N2/S2 and N6/S6, which were ejected only 1.3 hours apart.  This places an upper limit of 2.6 hours on the period, although the varying position angles of the intervening ejecta suggest that the true period is significantly shorter. The lower limit of order $\approx 1$ second is set by the lack of any blurring motion of the point source components over the timescale on which they are ejected ($>0.1$s; see Methods). Regardless, since the distribution of position angles for a precessing jet will peak at the two extremes, we can infer a precession cone half opening angle of $\sim18$\degree\ (Figure~2). 

Since V404 Cygni likely received a natal supernova kick\citemain{miller-jones09b}, a misalignment between the binary orbital plane and the black hole spin is expected. Plasma out of the black hole equatorial plane should then undergo Lense-Thirring precession\citemain{lense18}, potentially affected by torques from strong magnetic fields and associated jets\citemain{liska18}.
This phenomenon has been proposed to explain the low frequency quasi-periodic oscillations (QPOs) observed at sub-Eddington accretion rates in many X-ray binary systems\citemain{stella98,ingram16}. Regardless, both theoretical predictions and magnetohydrodynamic simulations\citemain{fragile07} of tilted disks have shown that
a sufficiently geometrically thick disk\citemain{papaloizou95} can precess as a solid body. To enable communication of the warp, the precession timescale must exceed the azimuthal sound crossing time of the disk.  The viscosity and magnetic fields should also be sufficiently low that the disk will not realign within a precession cycle\citemain{motta18}.

During its 2015 outburst, the X-ray behaviour of V404 Cygni could be explained by invoking a geometrically thick slim disk configuration\citemain{motta17}.  The mass accretion rate inferred from the peak X-ray luminosity implies a spherisation (outer) radius for the slim disk consistent with the maximum for solid body precession set by the viscous alignment timescale (see Methods).  This makes Lense-Thirring precession a plausible scenario for varying the disk orientation.  Precession of the inner slim disk would naturally result in precession of the jets, whether due to the magnetic field lines anchored in the precessing disk, or to realignment of spin-powered jets, either by powerful outflows from the inner disk\citemain{begelman06} or by the precessing slim disk itself\citemain{liska18}.

While the maximum radiative luminosity detected in the outburst was twice the Eddington luminosity\citemain{motta17}, super-Eddington accretion flows are known to drive powerful winds that can carry away a large fraction of the mass flowing in from the outer disk\citemain{poutanen07}, implying an outer accretion rate well above Eddington.  For moderate spins, mass inflow rates up to a few tens of times the Eddington accretion rate would imply precession periods\citemain{fragile07} of up to a few minutes and spherisation radii of a few tens to hundreds of gravitational radii (Extended Data Figure 5).  While such short periods would require the jet ejecta to be launched on timescales no longer than a few seconds, they would not require the jets to exceed the Eddington luminosity over the launching timescale (see Methods).  The precessing jets could also give rise to optical or infrared QPOs in the optically-thin synchrotron emission from the jet base.

A precessing accretion flow is also consistent with the marginal detections of short-lived low-frequency X-ray QPOs reported at 18\,mHz on June 22nd\citemain{huppenkothen17}. However, the link between the QPOs and the precessing disk is not clear and their short-lived nature would argue against long-term stable precession. In such a case, the changing mass accretion rate (and hence spherisation radius) would cause bursts of precession, subsequently damped by either disk alignment, or by changes in the sound speed\citemain{liska18,motta18}.  However, Figure~2 shows that the jet axis continues to vary over our full 2-week VLBA campaign. This suggests that precession continues with a relatively consistent cone opening angle, even if the precession timescale varies.

We have observed short-timescale changes in jet orientation from a black hole accreting near the Eddington rate, likely from a reservoir whose angular momentum is misaligned with the black hole spin. This spin-orbit misalignment in a low-mass X-ray binary suggests that the impact of black hole natal kicks can persist even after an evolutionary phase of accretion, and could therefore affect the observed gravitational waveforms\citemain{apostolatos94} during black hole merger events arising from the evolution of isolated binary systems.

Our findings are consistent with results from recent relativistic magnetohydrodynamic simulations, which demonstrated (albeit in the absence of radiation pressure) that the accretion flow and jets precess together, due to the combination of Lense-Thirring and pressure or magnetic torques from the inflow/outflow system\citemain{liska18}. The presence of a rapidly-precessing jet in a high-accretion rate source implies that varying jet inclination angles likely need to be accounted for when interpreting observations of systems such as ultraluminous X-ray sources\citemain{middleton18}, black hole-neutron star mergers\citemain{stone13}, gamma-ray bursts, tidal disruption events\citemain{lei13}, and rapidly-accreting quasars in the early Universe.

Kinetic feedback from precessing jets or uncollimated winds in AGN that distribute energy over large solid angles\citemain{falceta-goncalves10} has been invoked to prevent the onset of cooling flows in cool core clusters\citemain{vernaleo06} and to solve discrepancies between observed galactic properties and cosmological simulations\citemain{weinberger17}. For some low-luminosity AGN, which should host geometrically thick accretion flows, light curve periodicities and helical trajectories of jet components have been suggested as direct evidence of jet precession, typically attributed to the presence of a binary supermassive black hole\citemain{caproni04}.  However, Lense-Thirring precession can also match the observed timescales\citemain{nagai10,britzen18},
and might be expected in chaotic accretion scenarios.  Therefore, as demonstrated by our findings, precessing jets need not always signify binary black holes.

\bibliographystylemain{naturemag}
\bibliographymain{jmj_nature}

\begin{addendum}
 \item[Supplementary Information] is linked to the online version of the paper at www.nature.com/nature.

 \item The National Radio Astronomy Observatory is a facility of the National Science Foundation operated under cooperative agreement by Associated Universities, Inc.  JCAM-J is the recipient of an Australian Research Council Future Fellowship (FT140101082). AJT is supported by an Natural Sciences and Engineering Research Council of Canada (NSERC) Post-Graduate Doctoral Scholarship (PGSD2-490318-2016). AJT and GRS acknowledge support from NSERC Discovery Grants (RGPIN-402752-2011 \& RGPIN-06569-2016). MJM appreciates support via an STFC Ernest Rutherford Fellowship. DA acknowledges support from the Royal Society. GEA is the recipient of an Australian Research Council Discovery Early Career Researcher Award (project number DE180100346) funded by the Australian Government. TMB acknowledges financial contribution from the agreement ASI-INAF n.2017-14-H.0. PGJ acknowledges funding from the European Research Council under ERC Consolidator Grant agreement no 647208. SM and TDR acknowledge support from a Netherlands Organisation for Scientific Research (NWO) Veni Fellowship and Vici Grant, respectively. KPM acknowledges support from the Oxford Centre for Astrophysical Surveys, which is funded through the Hintze Family Charitable Foundation. KPM is currently a Jansky Fellow of the National Radio Astronomy Observatory. This work profited from discussions carried out during a meeting on multi-wavelength rapid variability organised at the International Space Science Institute (ISSI) Beijing by T.\ Belloni and D.\ Bhattacharya. The authors acknowledge the worldwide effort in observing this outburst, and the planning tools (created by Tom Marsh and coordinated by Christian Knigge) that enabled these observations.
 
 \item[Author contributions] JCAM-J wrote the manuscript with input from all authors. JCAM-J wrote the observing proposal BM421 with help from all authors.
 GRS wrote the observing proposal BS249 with help from JCAM-J, AJT, RPF, PGJ, GEA and KPM. JCAM-J designed and processed the VLBA observations. AJT performed the Monte Carlo modelling.  JCAM-J, AJT and GRS analysed the data. MJM led the development of the Lense-Thirring precession scenario, with help from S. Markoff.
 
 \item[Reprints] Reprints and permissions information is available at www.nature.com/reprints.
 
 \item[Competing Interests] The authors declare that they have no competing financial interests.
 
 \item[Correspondence] Correspondence and requests for materials
should be addressed to J.C.A.M.-J.~(email: james.miller-jones@curtin.edu.au).
\end{addendum}


\begin{figure}
\includegraphics[width=\textwidth]{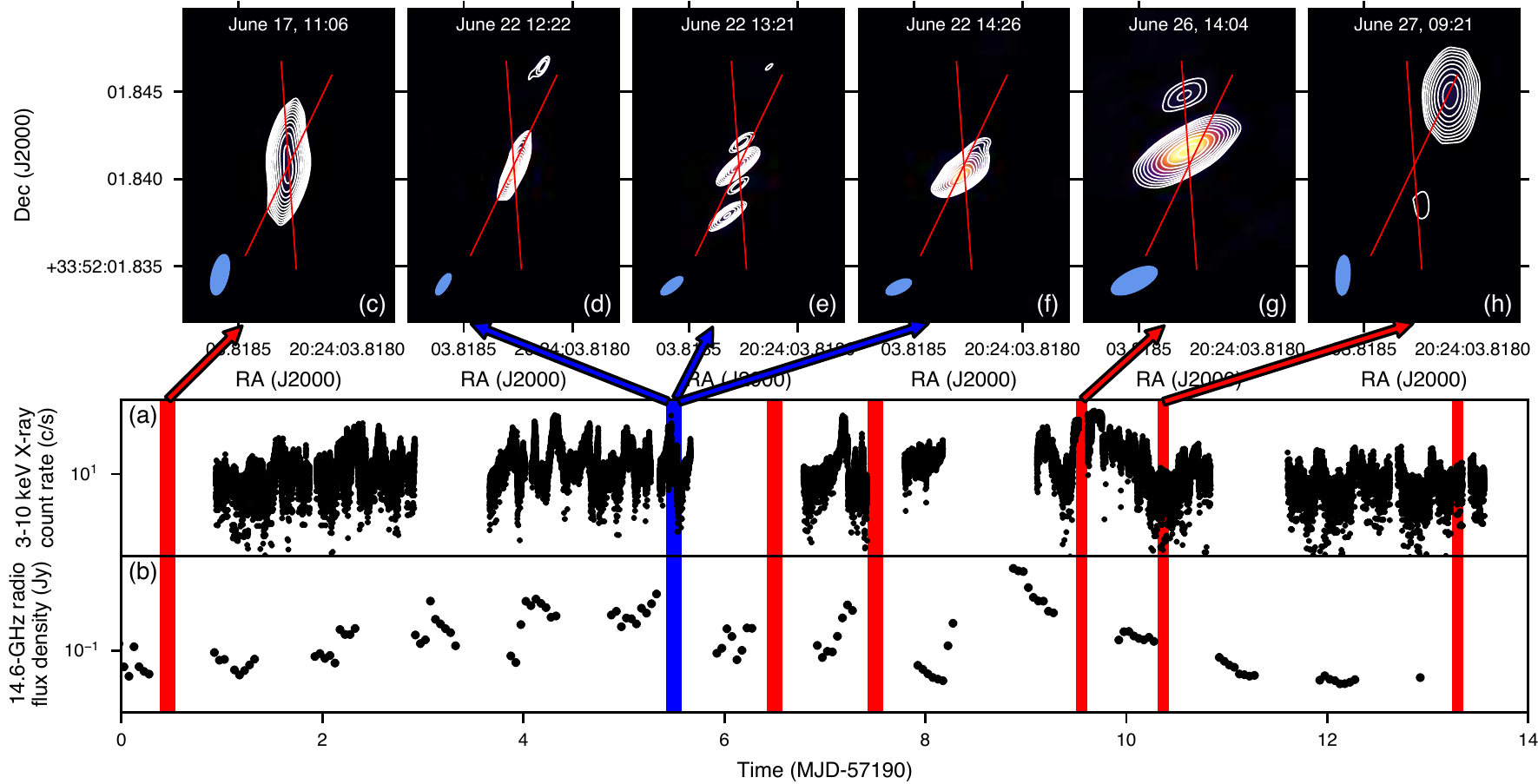}
\caption{{\bf VLBA monitoring of the radio jets during the 2015 outburst of V404 Cygni.} (a) 3--10 keV INTEGRAL X-ray count rate\protect\citemain{rodriguez15} over the brightest period of the outburst. (b) 14.6-GHz AMI radio light curve\protect\citemain{munoz-darias16}.  Red/blue shading show the times of our 8.4/15.4-GHz VLBA observations, respectively. (c-h) VLBA snapshot images, with observing dates as indicated. Blue ellipses show the synthesised beam shape, and red lines (centred on the radio core\protect\citemain{miller-jones09}, which is not detected on June 27th) show the measured range of position angles (Figure~2). The position angle of the ejecta changes over the course of the outburst, including over just a few hours on June 22nd.}
\label{fig:images}
\end{figure}

\begin{figure}
\includegraphics[width=0.99\textwidth]{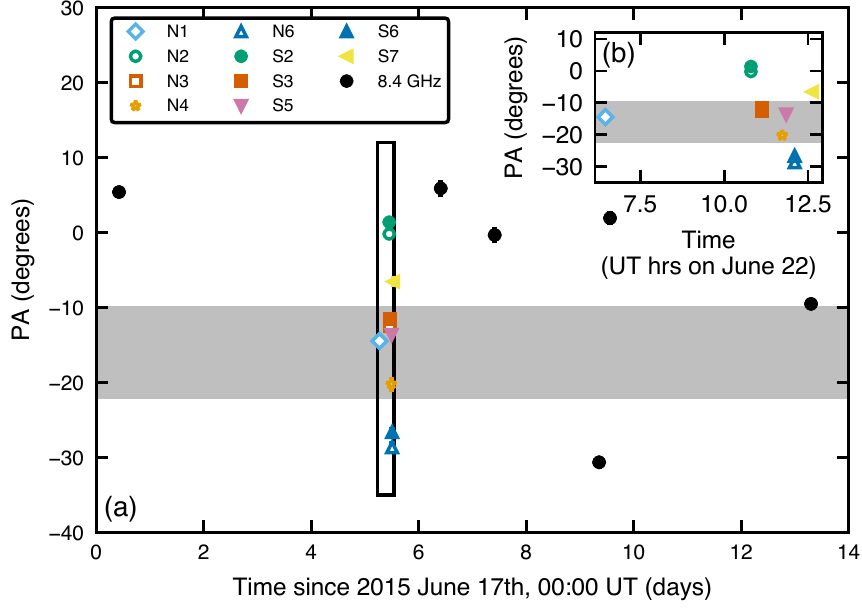}
\caption{
{\bf Jet component position angles.}  (a) Data from the full 14-day outburst period. Matched pairs of northern (N) and southern (S) components have the same colors. Uncertainties are shown at $1\sigma$. (b) Zoom-in on 15.4-GHz data from 2015 June 22nd, corresponding to the box in (a).  The true precession timescale is likely significantly shorter than the 2.6-hour upper limit inferred from pairs N2/S2 and N6/S6. The grey shaded region indicates the position angle of the quiescent jet inferred from the polarized radio emission during the 1989 outburst decay\protect\citemain{han92}, which is consistent with the central position angle that we measure in 2015.}
\label{fig:pa}
\end{figure}

\begin{figure}
\includegraphics[width=\textwidth]{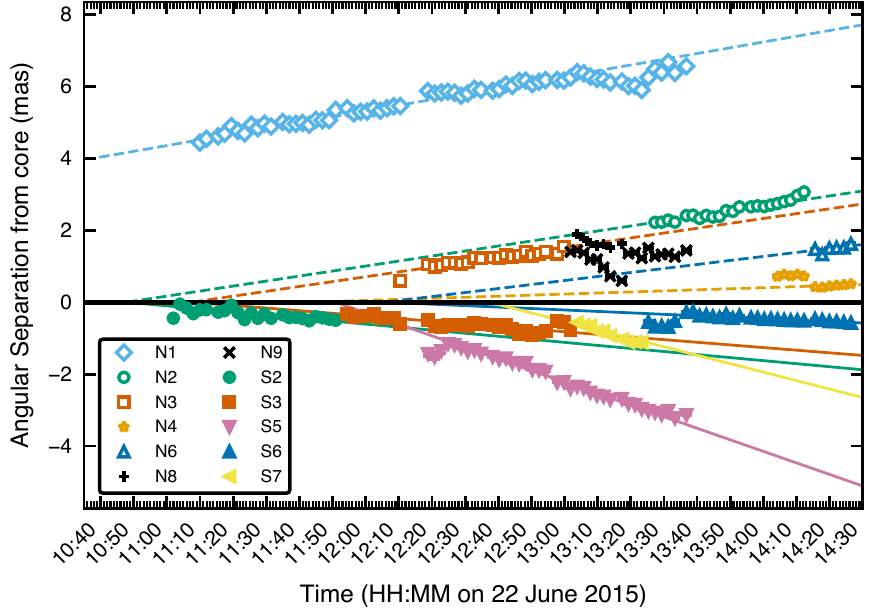}
\caption{{\bf Total angular separations from the core for all jet components on 2015 June 22nd.}  Positive and negative values denote displacements to the north and south of the core, respectively. Corresponding pairs of ejecta have matching colors and marker shapes. Uncertainties (typically smaller than the marker sizes) are shown at $1\sigma$. The best-fitting proper motions are shown as dashed (northern components; open markers) and solid (southern components; filled markers) lines.  All components except N8 and N9 move ballistically away from the core.  The fitted proper motions range from $4.3\pm0.1$ to $46.2\pm0.2$ mas\,day$^{-1}$ (N4 and S5, respectively).}
\label{fig:angseps}
\end{figure}

\begin{figure}
\includegraphics[width=\textwidth]{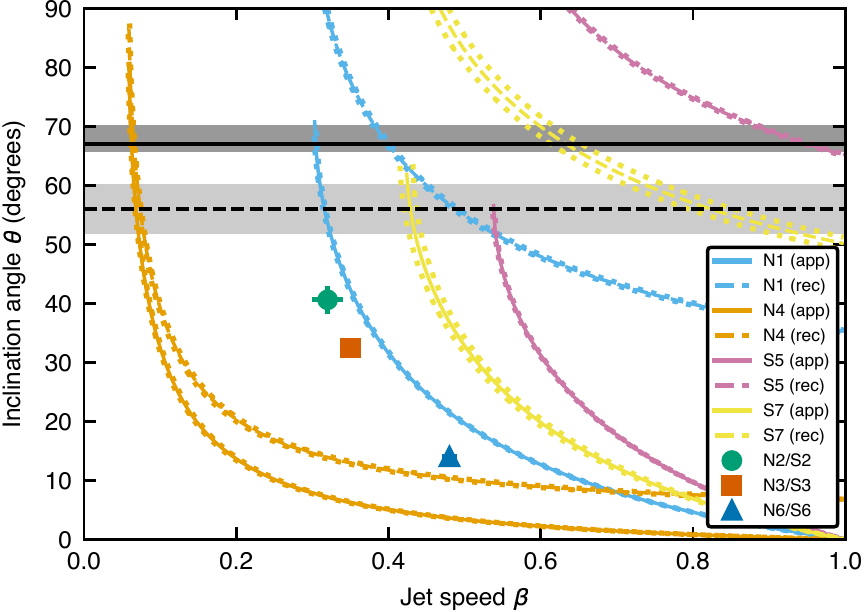} 
\caption{
{\bf Constraints on the jet speed and inclination angle to the line of sight.} Corresponding approaching and receding components on 2015 June 22nd allow the determination of both jet speed and inclination angle (individual points, with $1\sigma$ uncertainties). For unpaired components, the measured proper motion and source distance constrain only the product $\beta\cos\theta$, giving the plotted curves under the assumption of the components being either approaching (solid lines) or receding (dashed lines), with dotted lines showing $1\sigma$ uncertainties. The grey shading denotes the two published constraints on the orbital inclination angle\protect\citemain{shahbaz94,khargharia10}, which are inconsistent with the three ejecta pairs.}
\label{fig:betatheta}
\end{figure}

\newpage

\begin{methods}
V404 Cygni was observed over fifteen epochs with the VLBA, between 2015 June 17th and July 11th (Extended Data Table 1).

\subsection{Observations and data reduction.}
External gain calibration was performed using standard procedures within the Astronomical Image Processing System\citemethods{greisen03} (AIPS).  We used geodetic blocks to remove excess tropospheric delay and clock errors for all observations of duration $\geq 3$ hours.  Our phase reference calibrator was the bright (1.8\,Jy at 15\,GHz), nearby (16.6\,arcmin from V404 Cygni) extragalactic source J2025+3343\citemethods{ma98}.

The strong amplitude variability seen in both the VLBA data and the simultaneous VLA data from 2015 June 22nd\citemain{tetarenko17} violates a fundamental assumption of aperture synthesis.
We therefore broke the data down into short segments, within which the overall amplitude would not change by more than 10\%.
This equated to 103 scan-based (70-s) segments in the 15-GHz data from June 22nd, and two-scan (310-s) segments in the 8.4-GHz data from the other epochs.
The sparse {\it uv}-coverage in each individual segment meant that we could not reliably image complex structures.
We therefore minimised the number of degrees of freedom during deconvolution and self-calibration by performing {\it uv}-model fitting using the Difmap\citemethods{shepherd97} software package (v2.41), rather than the standard {\sc clean} algorithm.
With this approach, we found that the source could always be represented by a small number ($\leq 6$) of point source components.
To create the final images, we performed multiple rounds of phase-only self-calibration, and a final single round of amplitude and phase self-calibration (leaving noise-like residuals in all cases).

Since this version of Difmap did not provide uncertainties on the fitted model parameters, we used the Common Astronomy Software Application\citemethods{mcmullin07} (CASA; v4.7.2) to fit the self-calibrated data with the software tool UVMULTIFIT\citemethods{marti-vidal14}. We used the Difmap model fit results to define both the number of point sources used for each snapshot and the initial guesses for their positions and flux densities.

Given the sparse {\it uv}-sampling, we took additional steps to ensure the fidelity of our final images, taking guidance from previous time-resolved VLBI studies\citemethods{fomalont01}.
We examined each snapshot image to check for consistency between adjacent frames.
Only a small minority of frames showed inconsistent structure, and were therefore reprocessed using prior knowledge from the adjacent frames.
In a few cases, we imaged longer chunks of data (10--15\,min) to assess the fidelity of the structures with better {\it uv}-coverage.
As seen in Extended Data Figures 3--4, the positions and flux densities of our final set of components evolve smoothly with time (other than occasional jumps when a new component appears or a blend of two components separates sufficiently to become resolved).  This gives us confidence in the fidelity of our images.

\subsection{Markov Chain Monte Carlo analysis.}
Short-timescale tropospheric phase variations, particularly at 15.4\,GHz, coupled with the propensity of self-calibration to shift source positions by a small fraction of a synthesised beam combine to introduce low-level positional offsets between individual snapshots.  While these would be averaged out in longer data segments, they affected the fitted component positions in our snapshot images.
Furthermore, in snapshots made with fewer than 10 antennas (e.g.\ due to the source having set), poor {\it uv}-coverage made it hard to distinguish the true source position from the high sidelobes, and the initial peak position selected to start the model-fitting process dictated the astrometric registration of the final image.

To fit for the proper motions of the individual point source components on June 22nd, we first had to determine the positional offsets in each snapshot.
We assumed ballistic motion and constructed a set of linear equations with $k$ ejecta components and $i$ images, such that,
\begin{equation}
{\rm RA}_{ik}=\mu_{{\rm ra},k}(t_i-t_{{\rm ej},k})+{\rm J}_{{\rm ra},i}{\rm, \quad and}
\label{eq:ra}
\end{equation}
\begin{equation}
{\rm Dec}_{ik}=\mu_{{\rm dec},k}(t_i-t_{{\rm ej},k})+{\rm J}_{{\rm dec},i},
\label{eq:dec}
\end{equation}
where $\mu_{{\rm ra},k}$ and $\mu_{{\rm dec},k}$ represent the proper motions of the $k$th component, and $t_{{\rm ej},k}$ its ejection time.
The atmospheric jitter parameters ${\rm J}_{{\rm ra},i}$ and ${\rm J}_{{\rm dec},i}$ represent the offsets in position for the $i$th image, allowing us to correct the positional shifts.

With $k=10$ moving components (labelled by ejection time and direction of motion; see Extended Data Table~3), and $i=103$ images, we had 359 individual measurements in both right ascension and declination.  This translates to 20 linear equations, and 236 free parameters.
We took a Bayesian approach for parameter estimation, simultaneously solving equations (\ref{eq:ra}) and (\ref{eq:dec}) using a Markov-Chain Monte Carlo (MCMC) algorithm implemented with the \textsc{emcee} package\citemethods{foreman-mackey13}.
Prior distributions for all parameters are listed in Extended Data Table~4.
Lastly, due to the large number of rapidly-moving ejecta and the blending of components close to the core, it was occasionally difficult to distinguish between components.
We therefore assigned a confidence flag to each component for each image prior to the fitting (H = high, M = medium, L = low, and B = possible blended component) and weighted the data according to these flags (H=1, M=0.7, L=0.3, and B=0.1).

The best fitting results (Extended Data Table 3) were taken as the median of the posterior distributions from the converged MCMC solution, with the $1\sigma$ uncertainties reported as the range between the median and the 15th/85th percentile. Two components, N8 and N9, did not appear to move away from the core.
Given the faint nature of the components and the sparse {\it uv}-coverage, these could be artifacts arising from the difficulty of representing complex structures with a small number of unresolved point sources.

\subsection{Jet dynamics and Doppler boosting.}

From the similarities in ejection time and position angle, we identified three likely pairs of components (N2/S2, N3/S3, N6/S6).  In all cases, the proper motion of the northern component exceeded that of its southern counterpart, implying that the northern jets are approaching and the southern jets receding. This identification is supported by the first six epochs of our 8.4-GHz VLBA data, which all showed extensions to the north (see Figure~1), consistent with the northern components being both faster-moving and more Doppler-boosted.  Furthermore, only with approaching northern components do we get constraints on $\beta\cos\theta$ for the individual ejecta that are consistent with paired ejections (see Figure~4).

Assuming that our identification of pairs was correct, we then re-fit the proper motions of these three pairs, tying the ejection times of each component in a pair.
We use the results of these tied fits in Figures~2--4, and Extended Data Figures~1--3, and to calculate the jet physical parameters in Extended Data Table~5.  

Assuming intrinsically symmetric jets at a distance $d$, we can determine the jet speed and inclination angle from the proper motions of corresponding approaching and receding components via
\begin{align}
\mu_{\substack{{\rm app}\\{\rm rec}}} &= \frac{\beta\sin\theta}{1\mp\beta\cos\theta}\frac{c}{d},\label{eq:pms} \\
    \beta\cos\theta &= \frac{\mu_{\rm app}-\mu_{\rm rec}}{\mu_{\rm app}+\mu_{\rm rec}},\quad\quad{\rm and}\label{eq:bct}\\
    \tan\theta &= \frac{2d}{c}\frac{\mu_{\rm app}\,\mu_{\rm rec}}{\mu_{\rm app}-\mu_{\rm rec}}\label{eq:theta}.
\end{align}
With a known distance, equations (\ref{eq:bct}) and (\ref{eq:theta}) can be uniquely solved, allowing us to derive the jet Lorentz factor, $\Gamma = \left(1-\beta^2\right)^{-1/2}$ and the Doppler factors $\delta_{\rm app,rec} = \Gamma^{-1}\left(1\mp\beta\cos\theta\right)^{-1}$ (see Extended Data Table~5). For unpaired ejecta, we can only solve equation (\ref{eq:pms}) for $\beta\cos\theta$.

Given our estimated precession cone half-opening angle of $\approx18$\degree, the N2/S2 and N3/S3 pairs have inclinations consistent with being on the surface of a precession cone centred on the binary orbital angular momentum vector.
However, the N6/S6 pair has a very low inferred inclination of $14.0\pm0.8$\degree.
Either these two ejecta do not form a corresponding pair, or (more likely) the proper motion of N6 is affected by additional, unaccounted systematic uncertainties due to its slow motion and the short lever arm in time (it is based on only six points).
Thus this last pair should be treated as less reliable than the other two.  Even should N6 have been ejected slightly later, its observed angular separation suggests an ejection time prior to 13:40 UT, so our robust upper limit on the precession timescale remains a few hours.



\subsection{Mass accretion rate.}
The slim-disk geometry inferred from the X-ray emission implies an accretion rate at or above Eddington.  Further, the walls of the slim disk are likely to obscure the hottest inner regions of the accretion flow, implying an intrinsic luminosity higher than the maximum observed value of twice the Eddington luminosity ($2L_{\rm Edd}$)\citemain{motta17}.  Furthermore, a supercritical accretion disk is expected to launch a powerful outflow, which can expel a significant fraction of the infalling mass\citemain{poutanen07}.  Recent X-ray studies of ultraluminous X-ray sources have suggested that the wind kinetic power could be a few tens of times the bolometric luminosity (albeit reduced by the covering factor and solid angle of the wind)\citemethods{pinto16,pinto17}.  The mass accreted during the 2015 outburst was inferred to be a factor of three lower than the mass transferred from the secondary over the preceding 26-year quiescent period\citemethods{ziolkowski18}.  This was attributed to substantial wind mass loss, either from the outer disk\citemain{munoz-darias16} or from the inner regions\citemain{motta17}.  A total outer mass accretion rate of order ten times the Eddington rate would therefore be plausible, and would be sufficient to give rise to a precession period of order a minute (Extended Data Figure 5a).

The average bolometric luminosity over the outburst has been estimated as $\approx0.1L_{\rm Edd}$\citemethods{ziolkowski18}, suggesting that the outer mass accretion rate likely varied substantially.  This would alter both the spherisation radius $r_{\rm sph}$ and the precession period, and is consistent with the sporadic nature of the marginally-detected X-ray QPOs\citemain{huppenkothen17}.  This could suggest sporadic episodes of precession set by the changing mass accretion rate through the disk, rather than a long-term, stable, phase-coherent precession.   Assuming that the optical polarization (attributed to jet synchrotron emission) reflects the orientation of the jet axis, the slower inferred variation of the optical polarization position angle on June 24th ($4^{\circ}$ in $\sim30$\,min)\citemethods{shahbaz16} would support this scenario.

\subsection{Precession mechanisms.}
Various mechanisms have been put forward to explain X-ray binary jet precession.  In the slaved disk model (as applied to SS 433), tidal forces on the equatorial bulge of a misaligned early-type donor star cause the star to precess, thereby inducing the disk and jets to precess likewise\citemethods{roberts74}.  However, the predicted precession period\citemethods{hut81} for V404 Cygni is $\sim100$ times the 6.5-day orbital period, and cannot explain the observed changes in the jet axis.  Alternatively, massive outflows from a radiatively-warped, precessing outer disk could collimate and redirect the jets\citemain{begelman06}.  Existing treatments of radiatively-driven warping\citemethods{wijers99,ogilvie01} again predict precession periods significantly longer than the orbital period, although they were restricted to standard thin accretion disks ($H/R<\alpha$).  For more vertically-extended, super-critical disks, the outer disk (where the radiation warping instability acts most strongly) is shielded from the most luminous inner regions by the puffed up slim disk and the associated clumpy wind outflow, and radiation can be advected with the outflow, making radiative warps unlikely\citemain{middleton18}.

Resonances between the donor star orbit and the orbits of disk particles can also cause disk precession, giving rise to superhumps for systems with mass ratios $q\lesssim0.3$\citemethods{whitehurst91}.  However, the predicted periods are a few per cent longer than the orbital period, and again insufficient to explain the rapid changes we observed.  The tidal torque from the secondary is of order $10^{-9}$ times the Lense-Thirring torque at the spherisation radius, so cannot produce the required precession. Finally, since V404 Cygni is a dynamically-confirmed black hole, we can rule out precession driven by magnetic interactions between the compact object and the accretion disk\citemethods{mushtukov17}.

\subsection{Predicted precession period.}
The expected Lense-Thirring precession period for an inner super-critical accretion disk rotating as a solid body is\citemain{fragile07,middleton18}
\begin{equation}
P = \frac{\pi}{3a_{\ast}}\frac{GM}{c^3}r_{\rm sph}^3 \left[ \frac{1-\left(r_{\rm in}/r_{\rm sph}\right)^3}{\ln \left(r_{\rm sph}/r_{\rm in}\right)}\right],
\end{equation}
where $M$ is the black hole mass, $a_{\ast}$ is the dimensionless black hole spin $Jc/GM^2$ (with $J$ being the spin angular momentum), $G$ is the gravitational constant, and $r_{\rm in}$ and $r_{\rm sph}$ are the inner and outer radii of the slim disk (the latter being the spherisation radius), with all radii given in units of the gravitational radius $r_{\rm g} = GM/c^2$. 
We assume that $r_{\rm in}$ is located at the innermost stable circular orbit.  
Since the structure of the outer part of a supercritical disk is set by the angular momentum carried away by the disk wind, $r_{\rm sph}$ depends on the fraction of the radiation energy $\epsilon_{\rm w}$ used to launch the wind, as\citemain{poutanen07} 
\begin{equation}
    \frac{r_{\rm sph}/r_{\rm in}}{\dot{m}}\approx 1.34-0.4\epsilon_{\rm w}+0.1\epsilon_{\rm w}^2-(1.1-0.7\epsilon_{\rm w})\dot{m}^{-2/3},
\end{equation}
where $\dot{m}$ is the mass accretion rate in units of the Eddington rate.
The spin parameter of V404 Cygni was estimated\citemethods{walton17} as $a_{\ast}>0.92$, but without accounting for the slim disk geometry (which would require less light bending and hence a lower spin) and assumed the disk inclination to be that of the binary orbit, which our measurements show is not the case. The true spin could therefore be somewhat lower.
With a black hole mass of $12^{+3}_{-2}M_{\odot}$\citemain{shahbaz94}, we can then estimate the precession timescale of the slim disk for a given wind efficiency $\epsilon_{\rm w}=(1+L_{\rm rad}/L_{\rm wind})^{-1}$, where $L_{\rm rad}$ and $L_{\rm wind}$ are the radiative luminosity and wind power, respectively.

Based on the peak intrinsic luminosity\citemain{motta17}, and with a wind power fraction $\epsilon_{\rm w}$ of 0.25--0.5 (as estimated from relativistic magnetohydrodynamic simulations\citemethods{jiang14}), slim disk models imply $15<\dot{m}<150$\citemain{poutanen07}.  For moderate spins, we therefore predict precession timescales of order minutes and spherisation radii of tens to hundreds of $r_{\rm g}$ (see Extended Data Figure 5).
The predicted spherisation radii are consistent with the maximum radius expected for rigid precession\citemain{motta18}.  While the 18\,mHz QPO detected simultaneously with our observations (at 11:17 UT on June 22nd) was relatively low-significance at 3.5$\sigma$, it would imply a precession timescale of 56\,s. Given the uncertainty in mass accretion rate and black hole spin, this timescale is roughly consistent with these predictions.  
Since the maximum radius for rigid precession implied by the disk alignment criterion sets a spin and aspect-ratio dependent lower limit on the precession frequency\citemain{motta18}, then for an aspect ratio of $H/R=0.5$, this timescale would imply a spin of $a\lesssim0.3$.

\subsection{Jet energetics.}
The minimum amount of energy required to produce a given synchrotron luminosity is\citemethods{fender06}
\begin{equation}
    E_{\rm min} \approx 8\times 10^6 \eta^{4/7} \left(\frac{V}{{\rm cm}^3}\right)^{3/7} \left(\frac{\nu}{{\rm Hz}}\right)^{2/7} \left(\frac{L_{\nu}}{{\rm erg\,s}^{-1}{\rm\,Hz}^{-1}}\right)^{4/7}\,{\rm erg},
\end{equation}
where $\eta = (1+\beta)$ and $\beta$ is the ratio of energy in protons to that in the radiating electrons, $L_{\nu}$ is the monochromatic radio luminosity (given by $L_{\nu}=4\pi d^2 S_{\nu}$, where $S_{\nu}$ is the measured flux density), $\nu$ is the observing frequency and $V$ is the emitting volume.  We make the standard assumption that there is no energy in protons ($\eta=1$).  The brightest of our ejecta is knot S3, which at 12:07 UT has a flux density of 461\,mJy at 15.26\,GHz (Extended Data Figure 4), and is unresolved to the synthesised beam of $1.2\times0.4$\,mas$^{2}$.  Assuming a maximum knot radius of 0.4\,mas at 2.39\,kpc, we derive an upper limit on its minimum energy of $8\times10^{38}$\,erg.

While this knot would have been expanding adiabatically (with an expansion speed 0.01-0.15$c$\citemain{tetarenko17}), it never became significantly resolved to the VLBI beam, so should have been substantially smaller than 0.4\,mas at 12:07 UT. Hence the minimum energy is likely to be significantly lower than derived above.  On the other hand, if the magnetic field deviated significantly from equipartition, the energy could be somewhat higher than the minimum.

Should the precession period indeed be of order minutes, the knots would need to be launched over a timescale small enough that they were not significantly extended due to the precessional motion over the launching period.  This would argue for ejection on timescales no longer than a few seconds.  A lower limit on the timescale comes from the light crossing time of the jet acceleration zone, which was found to be 0.1 light seconds ($3\times10^9$\,cm)\citemethods{gandhi17}.  Alternatively, modelling the multi-frequency radio light curves gave fitted component radii of 0.6--1.3$\times10^{12}$\,cm at the peak of the sub-mm emission in each flare\citemain{tetarenko17}, corresponding to light crossing times of 20--40\,s. Since the sub-mm emission does not come from the jet base itself, the timescale of ejection would likely be significantly shorter.  In either case, our minimum energy synchrotron calculations above would not require the jets to exceed the Eddington luminosity.  However, even this would not be a hard limit given recent jet power constraints from ultraluminous X-ray sources\citemethods{pakull10,soria14}.

\subsection{Data availability}
The raw VLBA data are publicly available from the National Radio Astronomy Observatory archive (\texttt{https://archive.nrao.edu/archive/advquery.jsp}). All software packages used in our analysis (AIPS, Difmap, CASA, UVMULTIFIT, emcee) are publicly available. The final calibrated images and {\it uv} data are available from the corresponding author upon reasonable request.  The data underlying the figures are available as csv or xlsx files, and the measured positions and flux densities of all VLBA components from 2015 June 22nd are included with the MCMC fitting code (see below).

\subsection{Code availability}
The MCMC fitting code is available at \url{https://github.com/tetarenk/jet-jitter}.

\end{methods}


\bibliographystylemethods{naturemag}
\bibliographymethods{jmj_nature}


\noindent{\bf Extended Data}

\begin{figure}
\begin{center}
\includegraphics[width=\textwidth]{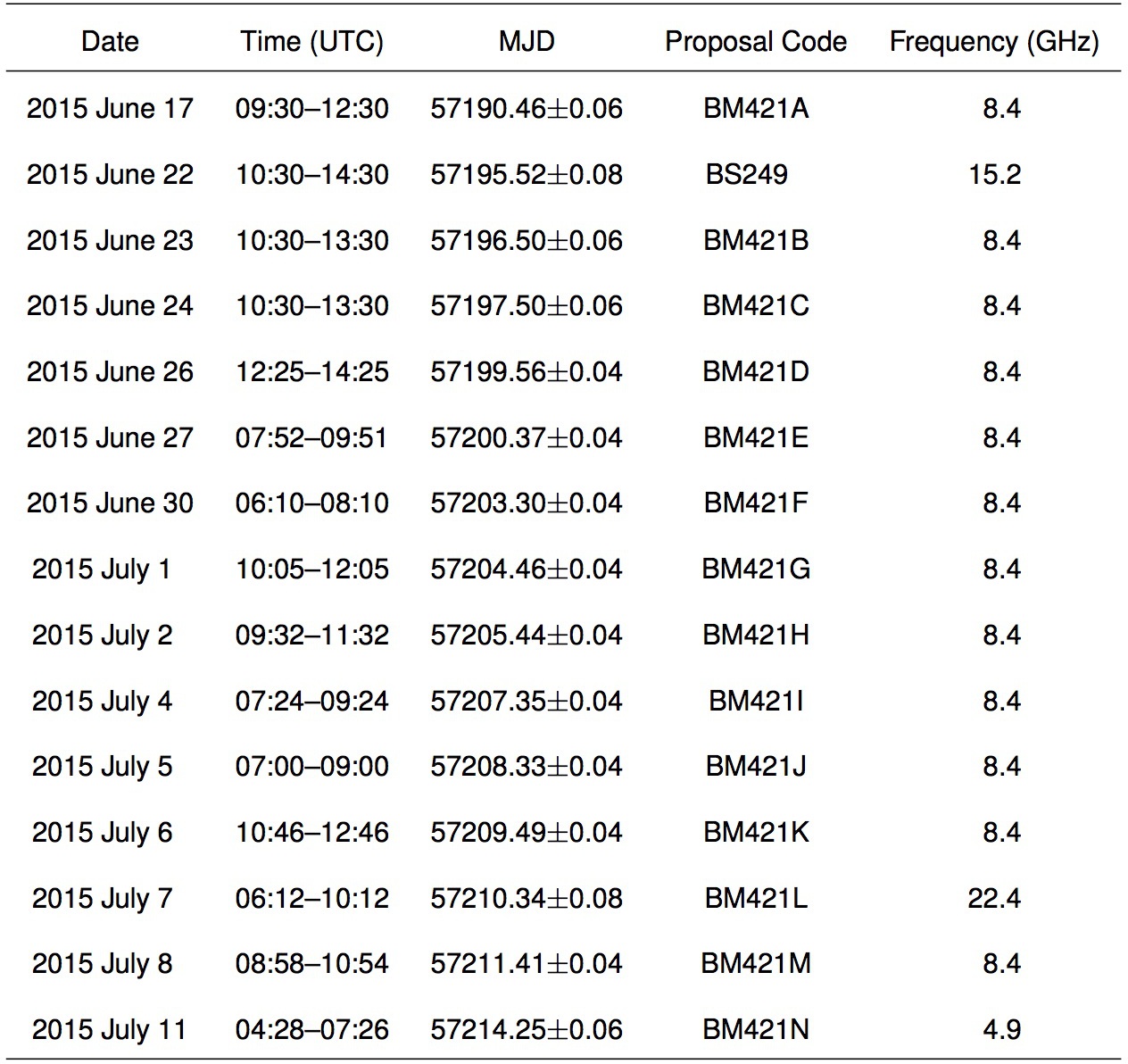}
\caption*{{\bf Extended Data Table 1: VLBA observing log for the June 2015 outburst of V404 Cygni.} Times denote the on-source time, and do not include the 30-min geodetic blocks at the start and end of the longer ($\geq 3$-hour) observations.}
\end{center}
\end{figure}

\begin{figure}
\begin{center}
\includegraphics[width=0.35\textwidth]{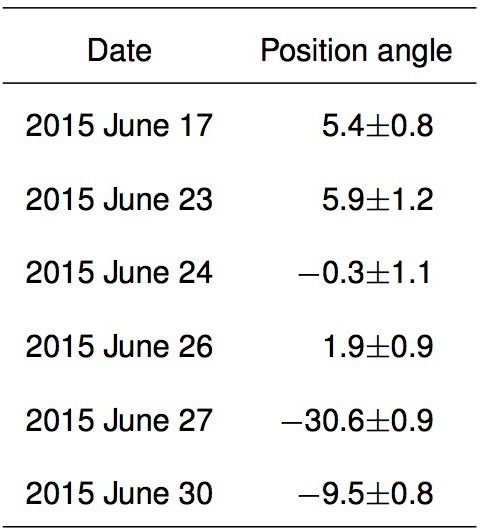}
\caption*{{\bf Extended Data Table 2: Measured position angles on the plane of the sky for the 8.4-GHz monitoring observations.} Position angles are measured in degrees east of north. The lower resolution at 8.4\,GHz meant that we only identified a single ejection event during each of these epochs, and the proper motions and ejection times could not always be well fit.}
\end{center}
\end{figure}

\begin{figure}
\begin{center}
\includegraphics[width=0.6\textwidth]{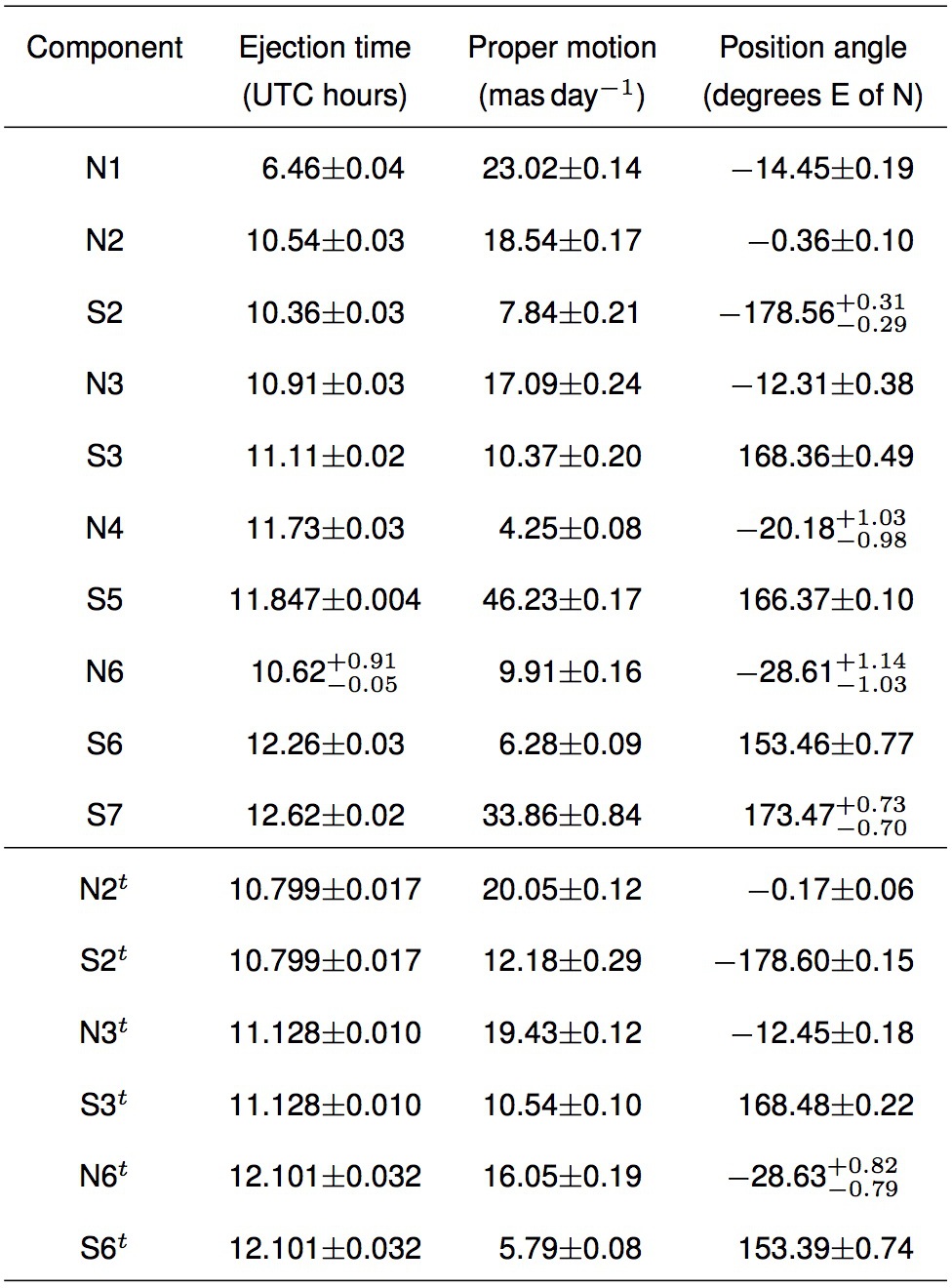}
\caption*{{\bf Extended Data Table 3: Measured component parameters for the 2015 June 22nd observations.}  N and S denote north- and south-moving ejecta, respectively.  From the similarities in their ejection times and position angles, we identify likely pairs of ejecta as N2/S2, N3/S3, and N6/S6.  Tying the ejection times of the two components of each pair gave the fits in the second section (denoted by the superscript $^{t}$). In cases where the parameters of the independent and tied fits differ significantly, the individual components either had relatively little data,
or little lever arm in angular separation.}
\end{center}
\end{figure}

\begin{figure}
\begin{center}
\includegraphics[width=\textwidth]{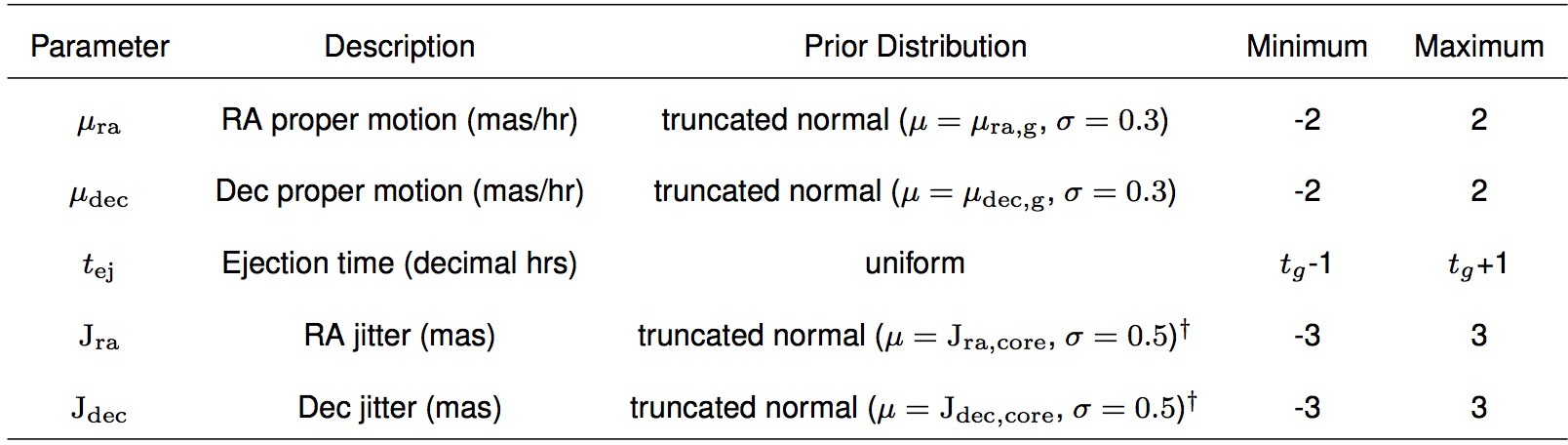}
\caption*{{\bf Extended Data Table 4: Prior distributions for atmospheric jitter correction model parameters}. Values with a subscript g represent the best initial guess for the parameter values. We use the offset positions (with respect to the center of the image) of the core jet component to represent the best initial guess for the jitter parameters $J_{\rm ra}$ and $J_{\rm dec}$, and to define their priors.}
\end{center}
\end{figure}

\begin{figure}
\begin{center}
\includegraphics[width=0.65\textwidth]{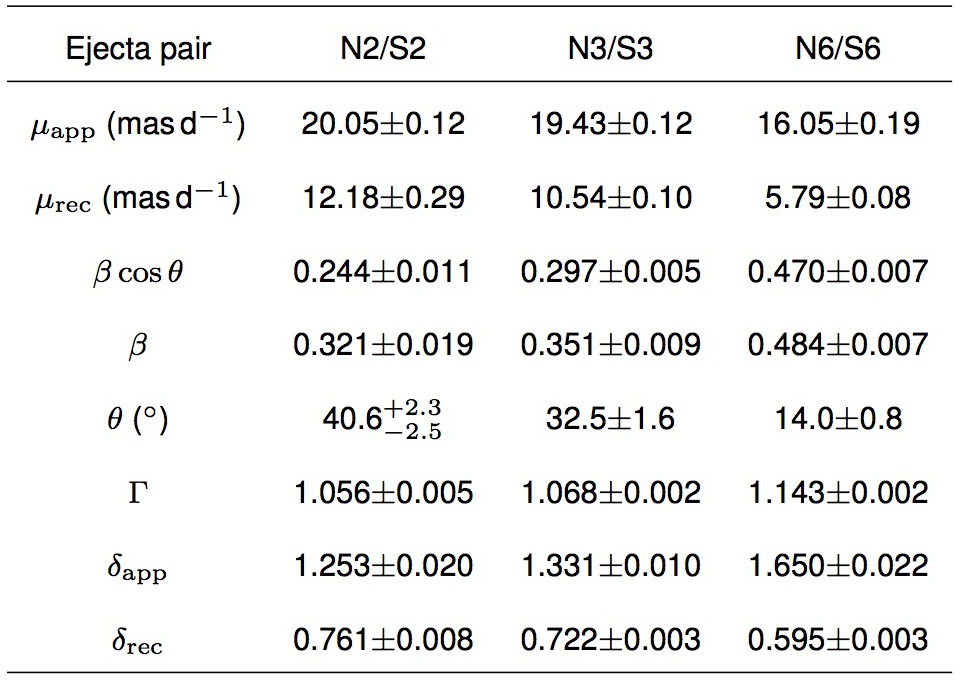}
\caption*{{\bf Extended Data Table 5: Inferred physical parameters from our identified paired ejecta from 2015 June 22nd.}  $\mu_{\rm app,rec}$ are the approaching and receding proper motions, $\beta$ is the jet speed as a fraction of the speed of light, $\theta$ is the inclination angle of the jet to the line of sight, $\Gamma$ is the jet bulk Lorentz factor, and $\delta_{\rm app,rec}$ are the approaching and receding jet Doppler factors.  In all cases the northern component is believed to be approaching and the southern component receding.}
\end{center}
\end{figure}

\begin{figure}
\begin{center}
\includegraphics[width=\textwidth]{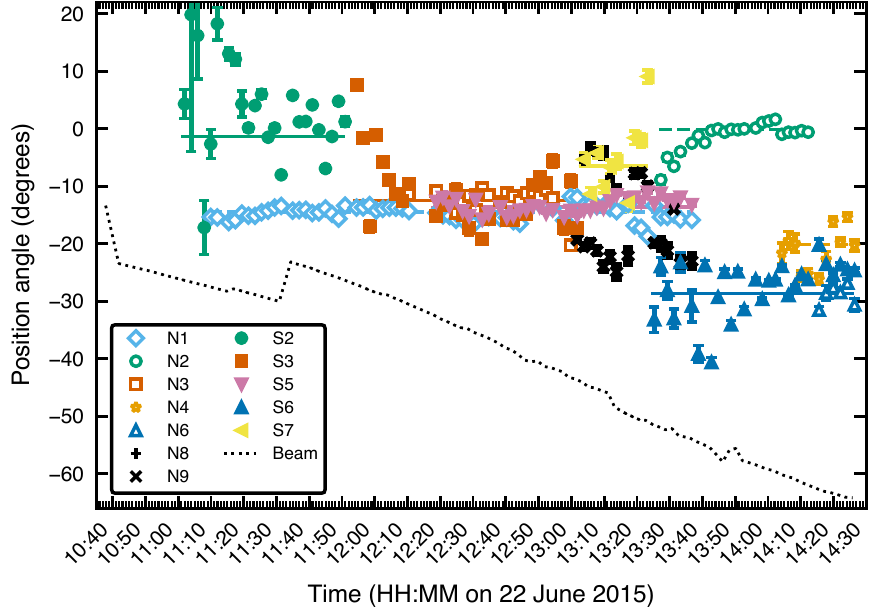}
\caption*{{\bf Extended Data Figure 1: Position angles of the jet components on June 22nd.} Angles are shown relative to the jitter-corrected centroid position, with $1\sigma$ uncertainties.  Corresponding pairs of components (N2/S2, N3/S3, N6/S6) are shown with matching colors and marker shapes.  The mean position angles of the components are shown as dashed (northern components) and solid (southern components) lines. Swings in position angle arise due to component blending as one gives way to another (e.g.\ S2/S3).  Dotted black line shows the orientation of the VLBA synthesised beam, which does not match the component position angles. Discrete jumps in beam orientation correspond to antennas entering or leaving the array.}
\end{center}
\end{figure}

\begin{figure}
\begin{center}
\includegraphics[width=\textwidth]{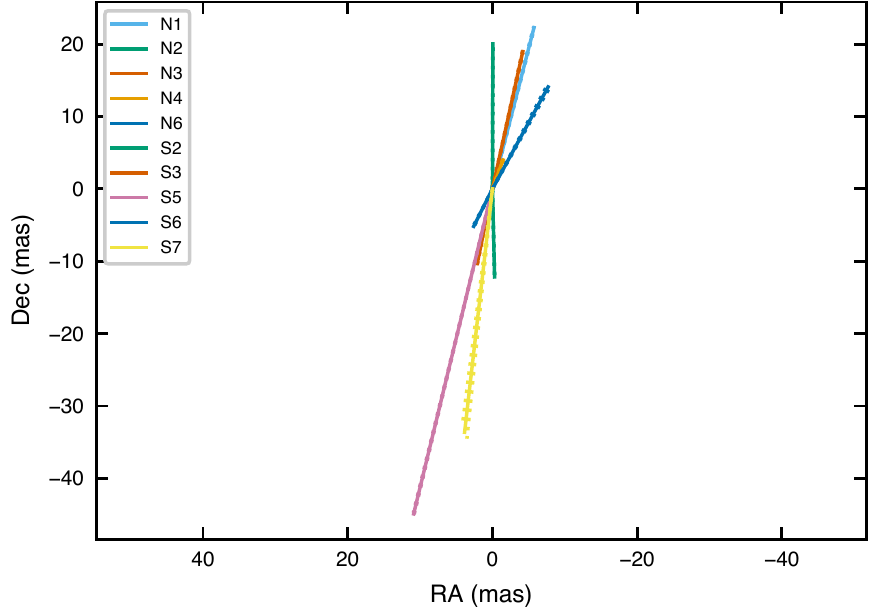}
\caption*{{\bf Extended Data Figure 2: Best-fitting proper motions of the different components on June 22nd.} Corresponding pairs of components (N2/S2, N3/S3, N6/S6) are shown with the same color.  Orientation shows the direction of motion, and length denotes the magnitude (distance travelled in one day). $1\sigma$ uncertainties are indicated by dotted lines (which, given the small uncertainties, merge into the solid lines).  The measured position angles range from $-0.2$ to $-28.6$\degree\ east of north (similar to that seen over the full outburst duration), providing a lower limit on the precession cone half-opening angle of $14.2$\degree.}
\end{center}
\end{figure}

\begin{figure}
\begin{center}
\includegraphics[width=\textwidth]{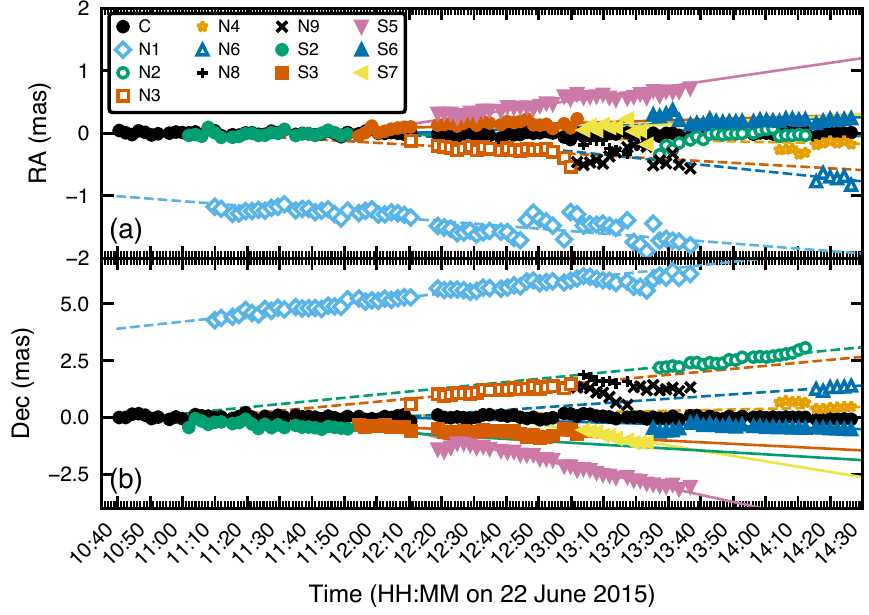}
\caption*{{\bf Extended Data Figure 3: Motions of the observed components on June 22nd.} Positions are corrected for atmospheric jitter, and shown in both Right Ascension (a) and Declination (b), with $1\sigma$ uncertainties (often smaller than the marker size). Corresponding pairs of ejecta have matching colors and marker shapes. The core is shown by filled black circles, and does not appear to move systematically over time. The best-fitting proper motions are shown as dashed (northern) and solid (southern) lines. The motion in Declination is larger than that in Right Ascension for all components. Other than N8 and N9, all components move ballistically away from the core.}
\end{center}
\end{figure}

\begin{figure}
\begin{center}
\includegraphics[width=\textwidth]{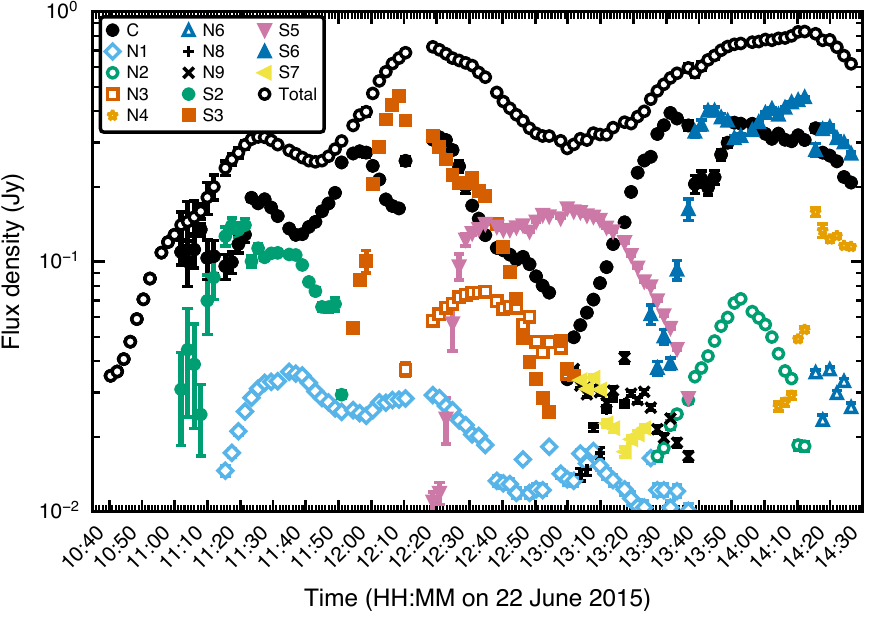}
\caption*{{\bf Extended Data Figure 4: Light curves of the individual components as a function of time on 2015 June 22nd.}  Corresponding pairs of ejecta have matching colors and marker shapes, with empty markers for northern components and filled markers for southern components.  Uncertainties are shown at $1\sigma$.  Top curve (empty black circles) indicates the integrated 15.4-GHz light curve (including the core source C).}
\end{center}
\end{figure}

\begin{figure}
\begin{center}
\includegraphics[width=\textwidth]{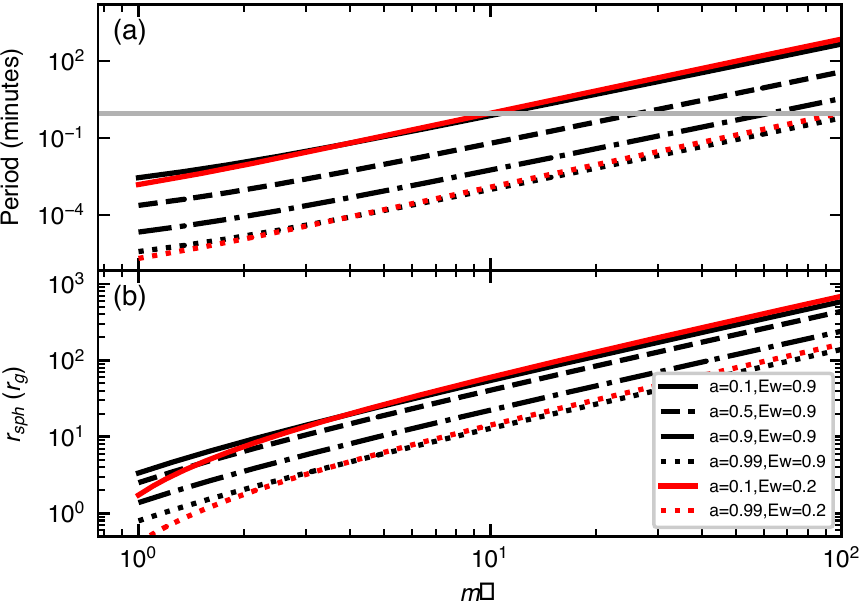}
\caption*{{\bf Extended Data Figure 5: Slim disk precession parameters.}(a) Calculated precession timescales and (b) spherisation radii (where the disk becomes geometrically thicker), as a function of Eddington-scaled mass accretion rate $\dot{m}$ and dimensionless spin parameter $a$.  The red lines illustrate the minimal impact of changing the fraction $\epsilon_{\rm w}$ of the accretion power used to launch the inner disk wind. The grey horizontal line in (a) shows the 18 mHz frequency of the most compelling X-ray QPO\citemain{huppenkothen17}.  For precession timescales of order minutes, we would need Eddington-scaled accretion rates of 10--100\,$\dot{m}_{\rm Edd}$ (depending on the black hole spin), corresponding to spherisation radii of 60--400\,$r_{\rm g}$.}
\end{center}
\end{figure}

\clearpage
\noindent
{\bf Supplementary Video: Movie showing the evolution of the jet morphology over four hours on 2015 June 22nd.}  Time (indicated in UT) has been sped up by a factor of 1000.  In the 103 separate snapshot images, we identify twelve separate components, together with a persistent core.   Ejected components appear to move ballistically outwards over time, with varying proper motions and position angles, implying precession of the jet axis.  Images have been corrected for atmospheric jitter (see Methods). Contours are at $\pm\sqrt{2}^n$ times the rms noise level of 3\,mJy\,beam$^{-1}$, where $n=3,4,5,...$. Top color bar is in units of mJy\,beam$^{-1}$.

\end{document}